# Three-dimensional hyperspectral imaging with optical microcombs

Stephan Amann,[1,2] Edoardo Vicentini,[2,3] Bingxin Xu,[1,2] Weiqiang Xie[4,*], Yang He,[5] Qiang Lin,[5] John Bowers,[4] Theodor W. Hänsch,[2,6] Kerry Vahala,[7] and Nathalie Picqué[1,2,8]

1. Max-Born Institute for Nonlinear Optics and Short Pulse Spectroscopy, 12489 Berlin, Germany
2. Max-Planck Institute of Quantum Optics, 85748 Garching, Germany
3. CIC nanoGUNE, 20018 Donostia-San Sebastian, Basque Country, Spain
4. University of California, Department of Electrical and Computer Engineering, Santa Barbara, CA 93106, USA
5. University of Rochester, Department of Electrical and Computer Engineering, Rochester, NY 14627, USA
6. Ludwig-Maximilian University of Munich, Faculty of Physics, 80799 Munich Germany
7. California Institute of Technology, T.J. Watson Laboratory of Applied Physics, Pasadena, CA 91125, USA
8. Humboldt University of Berlin, Department of Physics, 12489 Berlin, Germany
* Present address: Shanghai Jiao Tong University, School of Information and Electronic Engineering, Shanghai 200240, China

**Abstract**

Optical frequency combs have revolutionised time and frequency metrology [1, 2]. The advent of microresonator-based frequency combs ('microcombs' [3-5]) is set to lead to the miniaturisation of devices that are ideally suited to a wide range of applications, including microwave generation [6, 7], ranging [8-10], the precise calibration of astronomical spectrographs [11], neuromorphic computing [12, 13], high-bandwidth data communications[14], and quantum-optics [15, 16] platforms. Here, we introduce a new microcomb application for three-dimensional imaging. Our method can simultaneously determine the chemical identity and full three-dimensional geometry, including size, shape, depth, and spatial coordinates, of particulate matter ranging from micrometres to millimetres in size across nearly $10^5$ distinct image pixels. We demonstrate our technique using millimetre-sized plastic specimens (i.e. microplastics measuring less than 5 mm). We combine amplitude and phase analysis and achieve a throughput exceeding $1.2 \cdot 10^6$ pixels per second with micrometre-scale precision. Our method leverages the defining feature of microcombs — their large line spacing — to enable precise spectral diagnostics using microcombs with a repetition frequency of 1 THz. Our results suggest scalable operation over several million pixels and nanometre-scale axial resolution. Coupled with its high-speed, label-free and multiplexed capabilities, our approach provides a promising basis for environmental sensing, particularly for the real-time detection and characterisation of microplastic pollutants in aquatic ecosystems [17].

Three-dimensional (3D) hyperspectral imaging, which captures both spatial depth and spectral information at each pixel, represents a powerful yet underutilised technique for analysing complex materials. Environmental sensing, biomedical research, nanotechnologies or material science stand to benefit significantly from new approaches to 3D hypersectral imaging. Microcombs - broad spectra of narrow, equidistant lines- offer an ideal tool for this purpose. However, existing demonstrations have remained confined to pioneering but basic spectroscopy applications [18-20]. Despite the fundamental importance of sensing condensed matter, its exploration using frequency combs—especially microcombs—remains limited [19-22]. We take a new route to 3D hyperspectral imaging and present a transformative approach to material characterisation through the combination of large-linespacing microcombs and lensless digital holography. We leverage microcombs with very large line spacings, on the order of 1 THz, departing radically from the prevailing focus on microcombs with GHz-level spacing and open up new possibilities for precise, broadband interrogation of condensed matter systems.

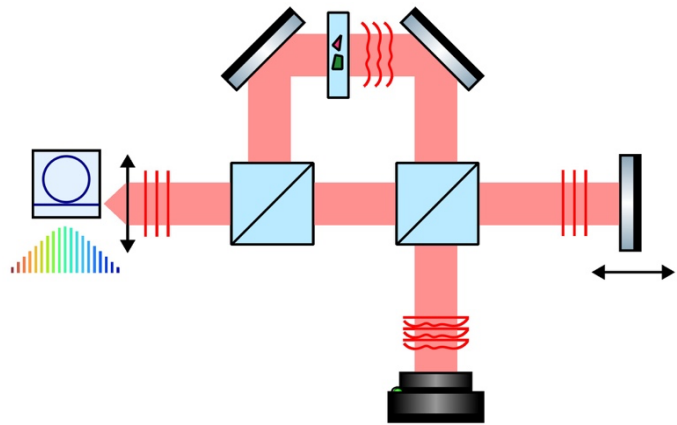

**Figure 1. Microcomb hyperspectral three-dimensional imaging setup.** The beam emitted by a microcomb generator split into two at a beamsplitter. One beam interrogates a spectrally-responsive three-dimensional object. The lightwaves scattered by the object are combined on a second beamsplitter with those of the second beam, which is reflected onto a moving mirror. Their spatio-temporal interference is recorded at a lensless camera sensor as a function of optical delay. Through subsequent harmonic analysis and holographic reconstructions, amplitude images and phase maps of the object are retrieved at any chosen depth.

**Principle of three-dimensional hyperspectral imaging with microcombs**

Our technique can uniquely identify the chemical composition, three-dimensional shape and position of particles across a wide field of view. It can also interferometrically measure the axial size and position of macroscopic particles beyond the diffraction limit. We use a spectrum of equidistant narrow lines with a large line spacing $f_{rep}$ (with $f_{rep}$>100 GHz) generated in a high-quality-factor microresonator driven by a continuous-wave laser, i.e. a microcomb. The use of a microcomb provides the large line spacing, a high power per comb line and a broad spectral span. The frequencies $f_n$ of the microcomb can be written as $f_n = f_{cw} + n f_{rep}$, where $f_{cw}$ is the frequency of the continuous-wave laser driving the microresonator and n is a small positive or negative integer. The beam of the microcomb is split into two beams (Fig. 1). One beam corresponds to that of the object microcomb, which illuminates the three-dimensional object and generates light scattered in the forward direction or in the backward direction (Supplementary Fig. 1). The second beam, that of the reference microcomb, is reflected on a moving mirror which is translated at a velocity $v$. The frequencies $f'_n$ of the lines of the reference microcomb are Doppler shifted: $f'_n = (f_{cw} + n f_{rep})(1 - 2v/c)$ where $c$ is the speed of light and the mirror is assumed to be moving away from the beamsplitter. The light scattered by the object is combined at a beamsplitter with the light wave of the reference beam. The waves interfere at a detector matrix, onto which an interference signal modulated in path difference – or optical delay- and in space is measured as a function of the optical delay between the two beams. This signal consists of amplitude and phase information encoded at all the audio frequencies $f'_n = 2v/c (f_{cw} + n f_{rep})$. The velocity of the moving mirror is chosen so that these down-converted frequencies are within the frequency bandwidth of the fast lensless camera sensor, which operates at rates faster than video rates. All the spatial and spectral information is simultaneously recorded on the sensor pixels, allowing throughputs in excess of $10^6$ resolution elements per second. We define a resolution element as an independent element of information in the lateral spatial plane and/or in the spectral domain. Here the number of spatial resolution elements is the number of pixels of the camera sensor and the lateral spatial resolution is the ratio of the product of the wavelength and the object-sensor distance to the sensor size. When the excursion of the moving mirror is longer than $c/(2f_{rep})$, the nominal spectral resolution of the interferometer is higher than $f_{rep}$ and therefore the individual microcomb lines can be isolated and as many spectral resolution elements as there are comb lines are obtained. Applying a complex Fourier transform to the interference signal in the optical-delay domain from each sensor pixel reveals the amplitude and phase of the interference pattern between the object wave and the reference wave as a function of the frequency. The frequency scale, in the audio range, can easily be converted back to the optical scale. Since we do not use any imaging optics, the spectral signal at each microcomb line across the detector pixels contains signatures of scattered waves from different parts of the object. This results in a complex-valued amplitude and phase hologram at a well-defined frequency.

Using known digital holography reconstruction techniques (Supplementary Information), the original complex-valued wavefront of the object wave can then be reconstructed at any focusing distance. Apart from the hyperspectral nature of our technique, which will be discussed below, our method allows to easily eliminate both the problematic twin image — the complex conjugate duplicate of the original image — and the zero-order background, which is the unmodulated part of the spatiotemporal interference pattern. The reconstructed amplitude images provide the hyperspectral response of the scene at any focusing distance and at a spectral resolution given by the microcomb line spacing $f_{rep}$. While the axial resolution in the amplitude images is typically two orders of magnitude coarser than the lateral resolution, a distinguishing feature of our technique is that we simultaneously acquire as many reconstructed phase maps as there are microcomb lines. While each of these phase maps enables highly precise length determination within a fraction – typically one tenth to one thousandth - of the optical wavelength $c/(2f_n)$, their determination is only modulo $c/(2f_n)$. By exploiting the entire hologram hypercube and combining the phase maps at different optical frequencies, the axial ambiguity range is extended to $c/(2f_{rep})$. Therefore, microcombs make unambiguous phase measurements possible over distances of hundreds of micrometers to several millimeters.

The use of a microcomb in this concept offers a variety of distinguishing features: the broad span, high power per comb line, and compatibility of combs with large line spacing and camera sensors (which have so far been significantly slower than single photodetectors) make the technique powerful. Additionally, using devices on photonics chips can lead to compact, miniaturised instruments. Therefore, we have overcome the limitations of dual-comb digital holography, as demonstrated using electro-optic frequency comb generators with a line spacing of 1 GHz [23]. Combining combs with narrow line spacing and camera sensors resulted in narrow spans that failed to provide sufficient spectral selectivity for condensed matter samples. The technique was only hyperspectral in the gas phase. Furthermore, the same narrow range rendered hierarchical phase unwrapping ineffective, since the iterative process of combining phase maps at different frequencies involved frequencies that were too close together to be useful.

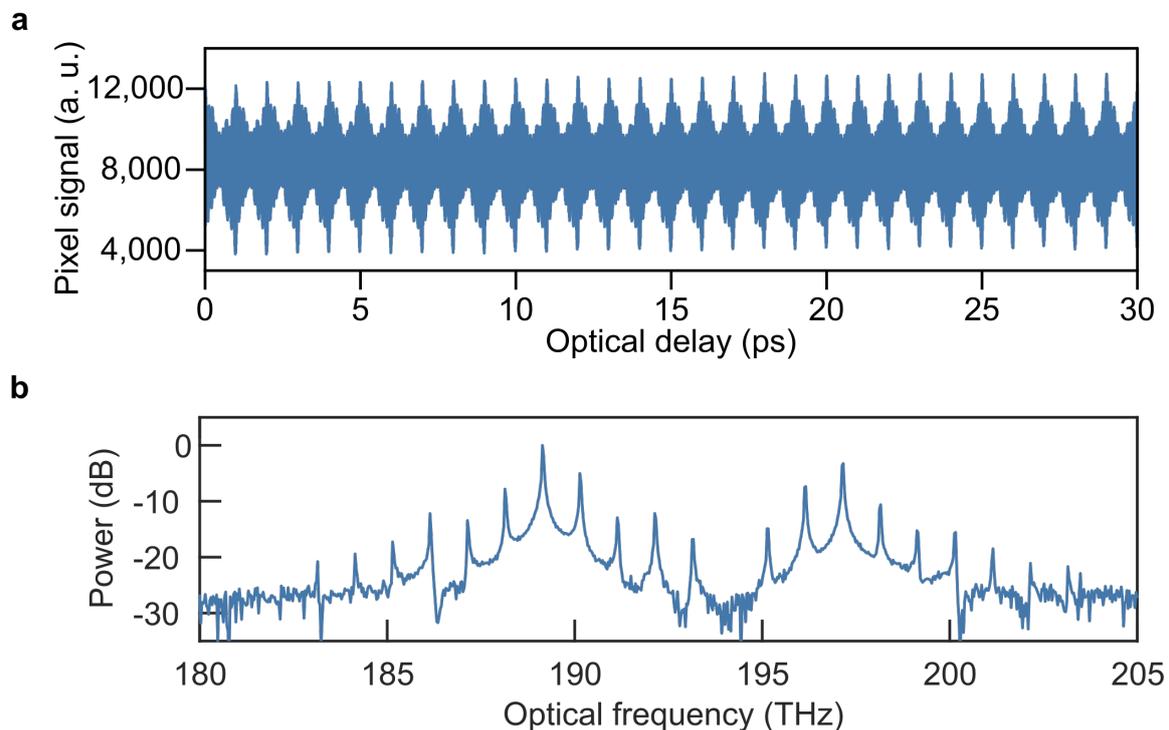

**Figure 2: Experimental interferogram and spectrum at one camera pixel**
a) Time-domain interference as a function of the optical delay, showing the reproducible 1-ps waveform period of the interfering microcomb waves. b) Microcomb spectrum, the Fourier transform of the interferogram in a. on a logarithmic y-scale. More than 20 comb lines, spaced by 1 THz, have a suitable signal-to-noise ratio for holographic measurements.

## Experimental demonstration of three-dimensional hyperspectral imaging with microcombs

We experimentally illustrate the potential of microcombs for three-dimensional hyperspectral imaging. A microcomb is generated by driving an AlGaAs microresonator [24] of a free-spectral range of 1 THz with a continuous-wave laser which emits at 193 THz (1552 nm). The beam of the microcomb is then analysed by an interferometer set up for interrogating an object in an inline forward- or backward-scattering configuration (Fig. 1, Supplementary Fig. 1). The light scattered by the object and that of the reference microcomb interfere on a fast InGaAs detector matrix. The lensless sensor has 256x320 pixels of a size of 30x30 $\mu m^2$ each and a frame rate of 300 frames per second.

In a first experiment, we apply our new concept for the characterization of small particles with a proof-of-principle demonstration of three-dimensional hyperspectral imaging of plastic samples. A compact instrument able to simultaneously chemically identify microplastic types and quantitatively describe their number, size, and surface properties is key to advancing the environmental monitoring and risk assessment of this global crisis, as well as analysing its impact on health and ecosystem disruption [25]. We fabricated an object comprising millimetre-sized particles of plastic polymer and small pieces of an optical bandpass filter, and analysed its transmission (Supplementary Information, Supplementary Fig. 1a). The object is placed in the beam path of the object microcomb. The spatio-temporal interference pattern, at each camera pixel, is recorded over a range of optical delays of 30 $10^{-12}$ s within 29 seconds (Fig.2a). The complex Fourier transform of each pixel reveals the phase and amplitude spectra of twenty microcomb lines, which are spaced by 1 THz and lie well above the noise floor.

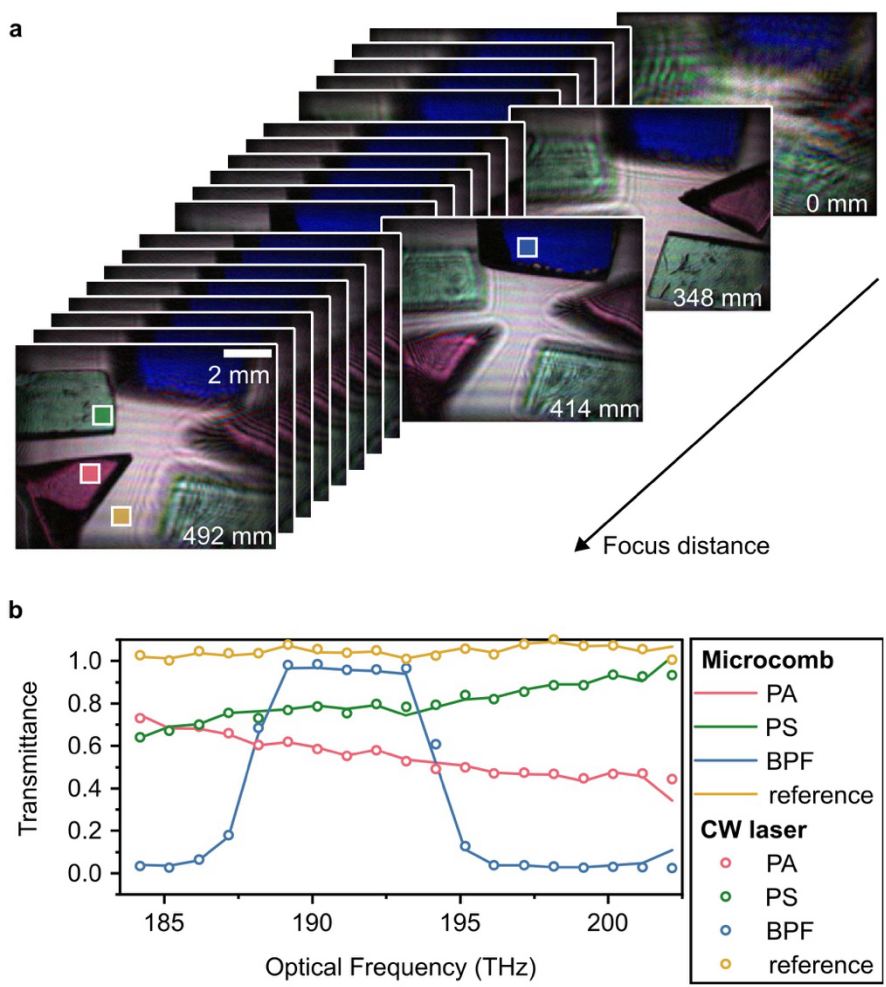

**Figure 3: Experimental three-dimensional hyperspectral imaging of plastic particles**
a) Microcomb holographic amplitude reconstruction using three different comb lines enables RGB-coded amplitude images to be used to reconstruct the scene at any chosen focus distance. Here, the focus distance ranges from 0 to 492 mm. Each of the 81,920 pixels in the amplitude cube corresponds to a spectrum with 1 THz resolution at a given spatial location, providing the spectral signature of species present in that part of the object. The boxes mark regions of 11×11 pixels that are averaged to produce. RGB coding is used for presentation but the reconstructed amplitude images and phase maps are obtained for the 20 microcomb lines shown in Fig. 2a.
b) Transmission spectra (lines) for 20 microcomb lines in the regions highlighted in (a). Each spectral sample in (b) is part of a pixel at a fixed frequency and a fixed focus distance in an amplitude and phase image cube. It provides the quantitative spatial distribution of a given species in the object with a distinguishable spectral signature at that frequency. The broad spectral span enables the unambiguous recognition of the spectral signatures of two different types of plastic (PA: polyamide and PS: polystyrene), as well as a bandpass filter (BPF). Validation spectra (circles) were obtained using a continuous-wave laser whose frequency was tuned stepwise to the position of the comb line.

These spectra span from 184 THz to 202 THz (1665–1462 nm) (Fig. 2b) and lead to 20 complex-valued holograms each comprising 81,920 pixels and corresponding to a specific optical frequency. The angular-spectrum method is used to reconstruct amplitude images and phase maps [26]. Reconstructed images at various focus distances are shown using a RGB colour scheme, where each colour corresponds to the amplitude image at a single microcomb line frequency (Fig. 3a). However, we have access to a much richer spectral characterisation, as we access twenty comb lines rather than three. Each pixel of the amplitude image corresponds to a spectrum spanning 20 THz (Fig. 3b). This broad range enables us to unambiguously identify and quantify the species that make up the object, even though the spectral features in the near-infrared region are broad and lack clear structure. To validate this, we reproduced the same experiment using a continuous-wave laser whose frequency was tuned in steps at the microcomb line frequencies. The two sets of data show excellent agreement, but the experiment with the continuous-wave laser, which includes its stepwise frequency tuning, takes over 1,000 seconds — 34 times longer than the duration of our microcomb experiment.

In a second experiment, we exploit the simultaneous access to amplitude and phase holograms, which enables to add, to the hyperspectral 3D imaging modality, axial resolution beyond the diffraction limit for microscopic objects, including those with steep depth variations. Microcombs enable efficient implementation of hierarchical phase unwrapping using simultaneously measured phase maps at different frequencies, as originally proposed using sequential measurements with distinct continuous-wave lasers. We measure the height profile of a 1-cent coin, analysed in backward scattering (Supplementary Information, Supplementary Fig. 1b) in a measurement of 0.96 seconds over 81,920 pixels. By combining the phase maps of 14 microcomb lines, we increase the non-ambiguity range from $c/(2f_n)$= 0.8 µm to $c/(2f_{rep})$= 150 µm. The resulting height map (Fig. 4) shows a standard deviation of 1.2µm in the center of the coin, possibly due to the surface roughness of the coin. In this measurement, a total of $1.14 \cdot 10^6$ independent spectral and spatial resolution elements of amplitude and phase are simultaneously measured within 0.96 seconds, leading to a throughput of $1.2 \cdot 10^6$ pixels s$^{-1}$.

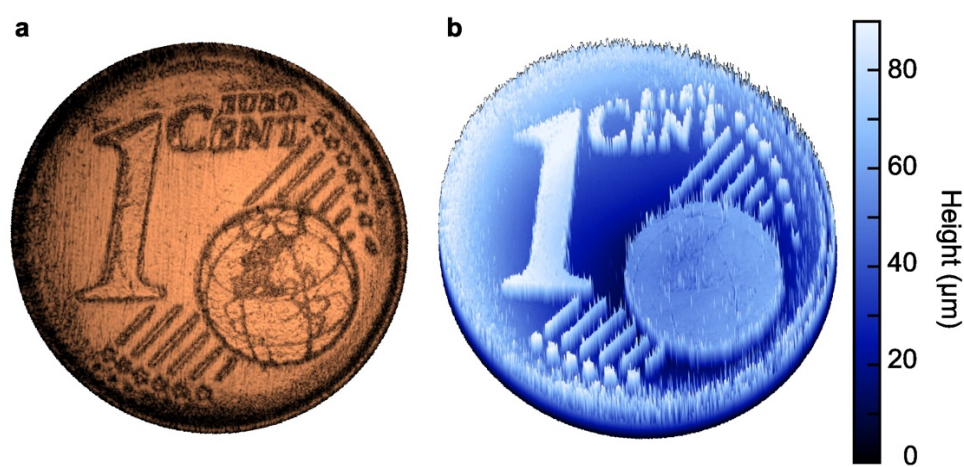

**Figure 4. High-throughput experimental height measurement using microcomb-based digital holography.** The object, a 1-cent coin, is measured in a backward-scattering configuration. a) The reconstructed amplitude of a single comb line shows a diffraction-limited image of the surface of the coin. The colour map only represents the greyscale value and does not reflect the coin's true colour. (b) By reconstructing and combining the phase information for 14 comb lines, a height map of the coin is obtained with a non-ambiguity range of 150 µm. (a) and (b) result from a single 0.96-second measurement at a sampling rate of $1.2 \cdot 10^6$ pixels s$^{-1}$

**Discussion and outlook**

We have devised a new technique that combines precise spectroscopic identification and quantification of chemical species in the condensed phase with three-dimensional imaging. Our technique achieves sub-diffraction limit axial resolution through phase imaging and diffraction-limited lateral resolution over a large field of view, using a lensless sensor. The current performance enables comprehensive characterisation of condensed matter with optical frequency combs. Despite the formidable technical challenges and exceptional promise for compact, rapid, multifunctional instruments, this area remains only sparsely explored. In contrast to previous demonstrations with microcombs that focused solely on spectroscopy, imaging or ranging, our approach uniquely delivers full-field, simultaneous acquisition of complex-valued wavefront data with spectral diagnostics, sub-wavelength relative depth resolution, and highly competitive throughput. The achieved acquisition rate of $1.2 \cdot 10^6$ pixels $s^{-1}$ compares favourably with coherent ranging and time-stretch lidar, while offering a substantial leap in spatial and spectral information content (Methods). Compared to the earlier dual-comb holography demonstration [23], we have reduced measurement times by over three orders of magnitude at an equivalent signal-to-noise ratio or improved the signal-to-noise ratio by approximately 65 within comparable acquisition times.

Our current frame rates already exceed video speeds by an order of magnitude, and improvements that are easily scalable — particularly through emerging CMOS sensors with rates approaching 500,000 frames per second at 640 x 480 pixels — could enable throughput enhancements of up to three orders of magnitude. This would reduce measurement times while pushing operation beyond the regime dominated by laser intensity noise. While the moving stage in our setup is not a limiting factor, entirely static interferometric configurations may offer greater compactness and be advantageous when ultra-fast cameras are deployed and may become key to flow-through analysis. Dual-microcomb systems are particularly attractive in this respect, provided that stabilisation of microcombs advances sufficiently. Furthermore, the demonstrated unambiguous identification of chemical species is poised to become significantly more powerful with increased spectral selectivity, attainable by targeting fundamental vibrational transitions, either via new comb spectroscopy combining coherent Raman effects [21] and photon-counting techniques [27] or by extending the technology into the mid-infrared. Operation in the ultraviolet range could further enhance lateral resolution. In all settings, specifically designed microcomb generators with a large line spacing, low noise and a flat-top intensity distribution will boost performance [15, 28-30]. Finally, improvements in axial precision will require stabilised soliton microcombs or real-time fluctuation tracking. Taken together, these avenues promise substantial advances in speed, resolution, and spectroscopic breadth, with the potential to redefine high-throughput, chemically specific 3D imaging across a broad range of condensed-phase systems.

The characterisation of particulate matter ranging in size from micrometres to millimetres has emerged as a topic of considerable scientific interest, presenting significant challenges in the field of instrumentation [25]. We propose new avenues of research with the long-term vision of developing a compact device capable of delivering a range of diagnostic capabilities that are currently provided by separate, specialised instruments. As underlying technologies continue to advance, such instrumentation could address critical issues in environmental and health monitoring. Applications could include detecting microplastics in food, water and air, measuring airborne asbestos fibres, characterising aerosols, dust or soot in industrial, combustion-derived or mining environments, monitoring volcanic ash and identifying pollen and spores for epidemiological studies. Other potential applications include assessing powder blending uniformity in pharmaceutical and industrial manufacturing, identifying chemical residues for explosive detection, and protecting cultural heritage by identifying pigments and materials, as well as monitoring degradation processes.

**Acknowledgments.** We thank Karl Linner for technical support, Jérémie Pilat and Lue Wu for useful discussions. This research has been supported by the Max-Planck Society, the Land of Berlin, the German federal government, the European Innovation Council Transition project MOLOKAI (Grant 101159440), the H2020 Marie Skodowska Curie Innovative Training Network Microcombsys (Grant 101119968), the Humboldt Foundation.

# Supplementary Information

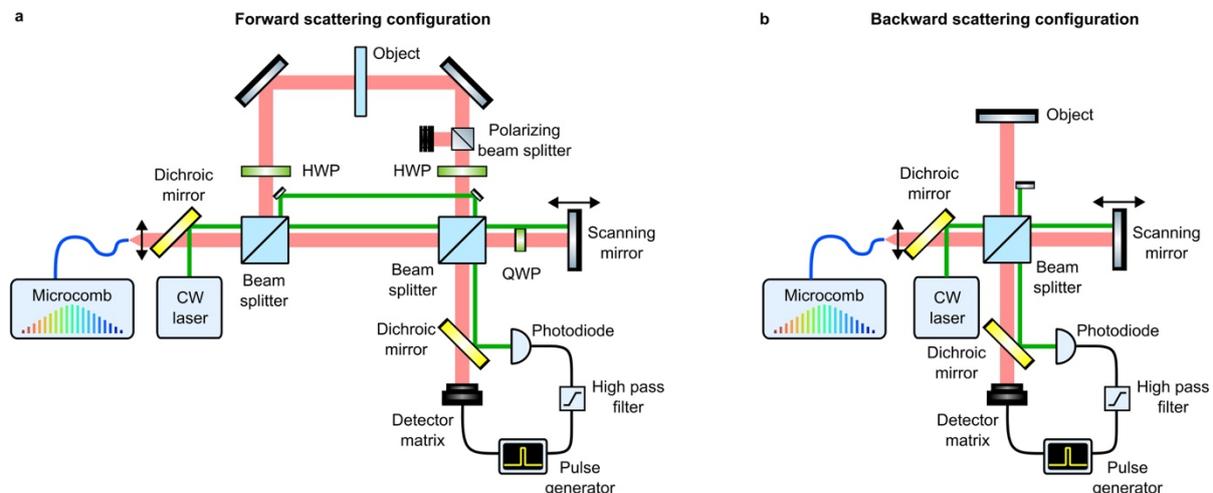

**Supplementary Figure 1. Detailed experimental schemes for the hyperspectral measurement setup.** The scattering of the object can be observed either in the forward direction (a) or in the backward direction (b). The waveplates and polarizing beam splitter in (a) are added in order to suppress parasitic reflection from the backside of the object. HWP: half waveplate. QWP: quarter waveplate. CW: continuous-wave

**Experimental setup for microcomb digital holography**

To generate a microcomb, we drive a microresonator with a high quality-factor on a photonic chip using a tunable laser system (Supplementary Fig. 2). An external-cavity diode laser emitting at 193.2 THz (1552 nm) is amplified to 130 mW in an erbium-doped fibre amplifier and coupled into the bus waveguide of an aluminium gallium arsenide (AlGaAs) microresonator using a lensed fibre. We utilise the AlGaAs-on-insulator platform [24], which is well-suited to low-threshold parametric oscillation due to the significant nonlinearity and high refractive index of AlGaAs. The microresonator employed, with a radius of 12.46 μm, has an intrinsic quality factor of 720,000. The waveguide has a width of 0.59 μm and a height of 0.4 μm. Using a calibrated, unbalanced fibre Mach–Zehnder interferometer with a free spectral range of 40 MHz, we measure the microresonator's dispersion parameters: $f_0$ = 193.6 THz, $D_1/(2\pi)$ = 1.002(48) THz, $D_2/(2\pi)$ = 335(16) MHz. The transmission spectrum of the microresonator shows strong mode-splitting of 600 MHz. The microcomb is generated by manually tuning the pump laser frequency into a resonance from the blue-detuned side at an on-chip pump power of approximately 30 mW. The light is coupled out of the chip using another lensed fibre, and remaining pump light is filtered out using a tunable fibre Bragg grating. The conversion efficiency is on the order of 3%. Two photodiodes, one positioned before and the other after the fibre Bragg grating, measure the power transmitted through the waveguides and that transferred into the microcomb.

Measured with an optical spectrum analyser, the microcomb spans more than 20 THz (160 nm), with 21 lines in a dynamic range of 20 dB (Supplementary Fig. 3). We perform the measurements in a microcomb state that is not phase-locked. This state was chosen for its low noise in the detection bandwidth, and for the slower fall-off of the spectral intensity profile of the microcomb compared to a solitonic sech², which is key to accessing a large number of comb lines with a sufficient signal-to-noise ratio. A fibre polarization controller and a polarizer are used to adjust the microcomb power in the holography experiment.

The microcomb generator acts as the light source in a digital holography setup that employs a scanning two-beam interferometer (Fig. 1). The interferometer serves as a spectrometer, resolving the microcomb lines and eliminating the unwanted twin image and zeroth-order light that obscure the reconstruction of the object. The interferometer's optical layout depends on whether the scattering of the object is collected in the forward or backward direction (Supplementary Fig. 1). The object is placed in the fixed arm of the interferometer. The reference beam is provided by the moving arm of the interferometer, which has a plane mirror mounted on a motorised linear translation stage. This generates a microcomb that is Doppler shifted with respect to the object beam. The two beams are combined on a beam splitter and interfere at the camera sensor. The InGaAs sensor has 256 x 320 pixels, each measuring 30 x 30 µm², and an acquisition rate of 300 frames per second. Its dynamic range is 14 bits. Each frame is integrated during 60 µs. Data acquisition at even optical delay intervals is triggered by the zero crossing of the interference pattern, which is provided by a continuous wave, single frequency laser operating at 288 THz (1040 nm) and following a neighbouring beam path in the interferometer. The interferogram hypercube is streamed to a personal computer via a frame grabber and stored as a binary file on a hard drive.

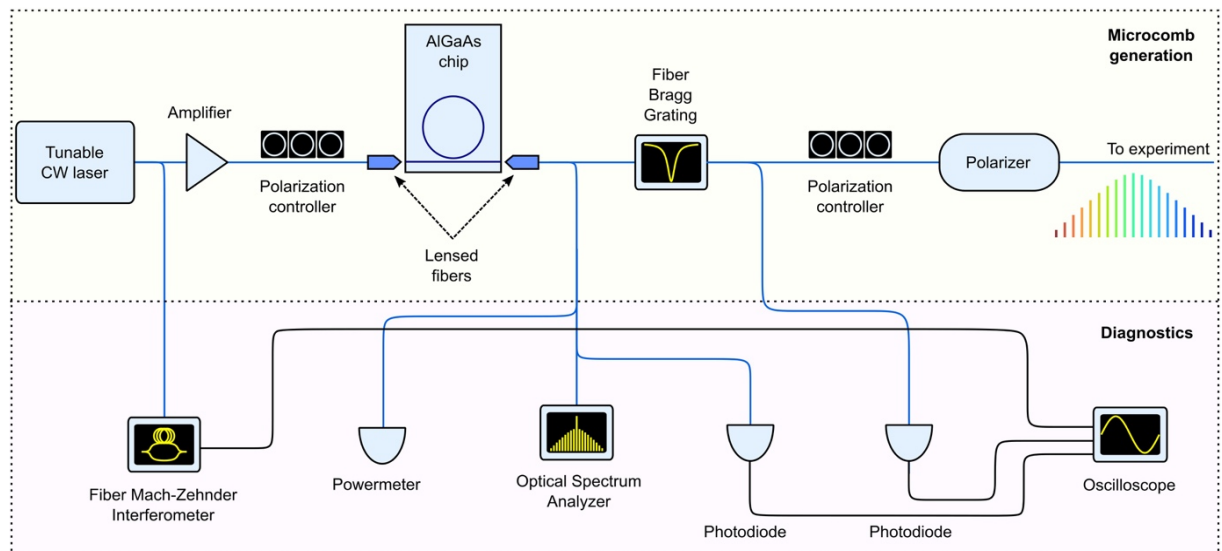

**Supplementary Figure 2: Experimental setup for the generation and analysis of the microcomb.** A tunable continuous-wave laser is amplified and coupled into the bus waveguide of an AlGaAs on-insulator high-quality-factor microresonator using a lensed fibre. A microcomb is then coupled out. The residual driving light is filtered out using a fibre Bragg grating and the microcomb light is directed towards the three-dimensional hyperspectral imaging setup. A calibrated fibre Mach–Zehnder interferometer is used for frequency calibration when measuring the microresonator's dispersion. An optical spectrum analyser measures the generated microcomb. Calibrated photodiodes before and after the fibre Bragg grating enable analysis of the full transmission through the device and that transferred into the microcomb.

For each pixel, the time-domain interferogram is zero-filled six times and then undergoes a Fourier transform to reveal the complex-valued spectrum (amplitude and phase). The microcomb line positions are then retrieved from the spectrum at a single pixel using peak-finding algorithms. The spectral amplitude and phase at these frequencies are then retrieved from the spectrum of each pixel, creating a complex-valued hologram hypercube comprising 320 x 256 x N amplitude and phase data points, where N is the number of microcomb lines. An amplitude and phase hologram at a microcomb line frequency comprises 320 x 256 pixels at that frequency. Each hologram is zero-filled by 600 pixels and apodised with a cosine filter to prevent artefacts from the image borders during reconstruction. Through the use of reconstruction algorithms based on the theory of Fresnel–Kirchhoff diffraction [26], a hologram is multiplied by a predefined reference wave, and the angular-spectrum method computes backpropagation at any chosen axial distance, leading to amplitude images and phase maps.

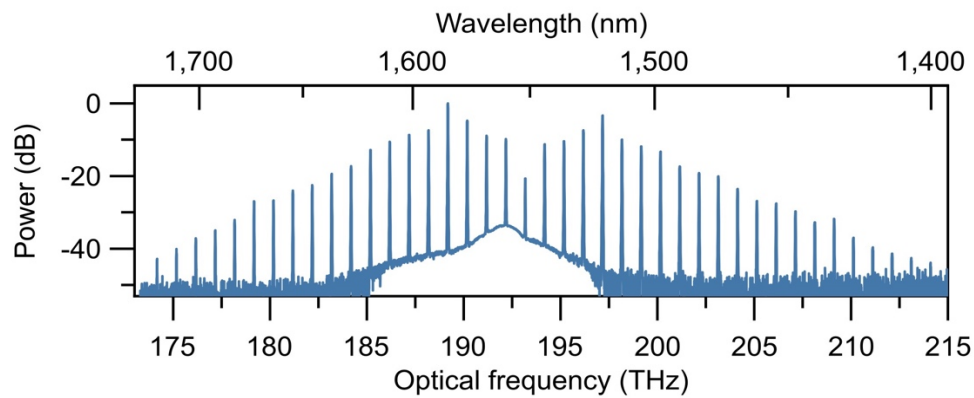

**Supplementary Figure 3. Microcomb characterization.** Microcomb measured on an optical spectrum analyzer, spanning over 20 THz (160 nm) at a -20dB bandwidth.

**Recording conditions and data processing in microcomb digital holography**
For the hyperspectral, three-dimensional imaging measurements of the plastic samples presented in Fig. 3, the objects were prepared as follows: small pieces, around 3 mm in size and of various shapes, were cut from plastic sheets. Their edges were smoothed with sandpaper. The polystyrene (PS) samples were cut into rectangles from a 1.2 mm thick sheet. The polyamide (PA) samples consisted of a 1 mm thick Nylon 6,6 sheet cut into triangles. A small drop of glue was placed on one corner of each sample (not visible in the image) to fix them onto two microscope slides, which were then positioned in different locations within the interferometer's fixed object arm. Additionally, a piece of an optical bandpass filter with a centre frequency of 190 THz (1575 nm) and a full width at half maximum of 6 THz (50 nm) was employed. The optical filter is a standard thin-film dielectric coating on a fused silica substrate, which was broken into small pieces. Parasitic reflection from the back of the slides is suppressed by the addition of waveplates and a polarising beam splitter. During recording, the mirror of the interferometer is scanned over a distance of 4.5 mm at a velocity of 0.16 mm/s. This corresponds to a maximum optical delay of 30 $10^{-12}$ s (Fig. 2a), meaning the interferometer has an unapodised resolution of 33 GHz. The measurement time is 29 seconds. It takes 60 seconds to compute the Fourier transform of the acquired interferogram hypercube, which comprises 81,920 interferograms of 8,700 time-samples each, on a standard consumer personal computer. The resulting spectral hypercube shows twenty microcomb lines that are sufficiently above the noise floor (Fig. 2b). The signal-to-noise ratio, defined as the height of a comb line relative to the noise floor, reaches 33 dB (2,000 on a linear scale) for the strongest microcomb line. Since the microcomb line spacing is 1 THz and the microcomb lines are clearly resolved by the interferometer, the spectral resolution in the hyperspectral three-dimensional imaging experiment is also 1 THz. Transmittance spectra of the same species at different lateral and axial locations are in good agreement (Supplementary Fig. 4). In the validation measurement shown in Fig. 3b, the frequency of a continuous-wave, single-frequency laser is tuned to the positions of the microcomb lines. A hologram is measured at each of 19 microcomb line frequencies using the continuous-wave laser and the digital holography interferometric setup, under conditions that are otherwise identical to those used with the microcomb generator. In Fig. 3b, each spectrum is reconstructed in the focal plane of the particle that is highlighted in Fig. 3a. For each particle, the spectra from an area of 11 x 11 pixels are averaged. The transmittance spectra in Fig. 3b are the ratio of these spectra to the reference spectrum, which is set to 100% transmission. The reference spectrum is taken from a region of 11 x 11 pixels in the centre of the image that does not contain any objects.

For amplitude and phase holographic measurements of a coin (Fig. 4), the interferometer mirror is scanned over a total distance of 0.6 mm at a velocity of 0.62 mm/s. This results in a maximum optical delay of 2 $10^{-12}$ s, enabling the interferometer to achieve a spectral resolution of 250 GHz. The measurement time is 0.96 seconds. The complex Fourier transform of the interferogram hypercube, comprising 81,920 interferograms of 288 time-samples each is computed within 2.5 seconds. Fourteen microcomb lines with a line spacing of $f_{rep}$ = 1 THz are sufficiently above the noise floor. The strongest microcomb line has a signal-to-noise ratio of 18 dB (63 on a linear scale). In Fig. 4b, the amplitude image and the height map are shown to be reconstructed in the focal plane of the coin. For one microcomb line, the phase distribution over the spatial interference pattern is measured only modulo 2π, just as it is for a single-frequency continuous-wave laser. Consequently, each measurement is made modulo half the wavelength. This makes the technique suitable only for microscopic objects smaller than half the wavelength (e.g. smaller than 1 μm) or for objects where the optical path difference between adjacent pixels is assumed to be less than half the wavelength and the surface under test is smooth. This technique is not suitable for characterising wavefronts with steep variations. Microcombs allow hierarchical phase reconstruction to be implemented efficiently, as was initially proposed for distinct serial measurements with different continuous lasers [31]. With a microcomb, all the phase maps at comb line frequencies can be retrieved simultaneously from a single recording. By subtracting phase maps between neighbouring comb lines with a high signal-to-noise ratio and averaging them, the unambiguous measurement range extends to $c/(2f_{rep})$, although this increases the noise. Multiple frequency combinations corresponding to multiples of the comb line spacing are then included up to the optical frequency of a comb line. Phase information is then propagated iteratively, enabling highly accurate, denoised quantitative phase maps to be retrieved while maintaining the benefit of a large unambiguous range, finally leading to a height map. In Fig. 4b, the non-ambiguity range of the height measurement, which is equal to $c/(2f_{rep})$, amounts to 150 μm. The standard deviation of the reconstructed height map over an area of 60x60 pixels in the center of the coin, corrected by a second-order polynomial fit is 1.2 μm. This sets an upper limit on the precision of the measurement, although it is likely due to the surface quality of the coin. Future work will involve a determination with calibrated gage blocks and a careful assessment of the uncertainties and systematic effects in our experiments.

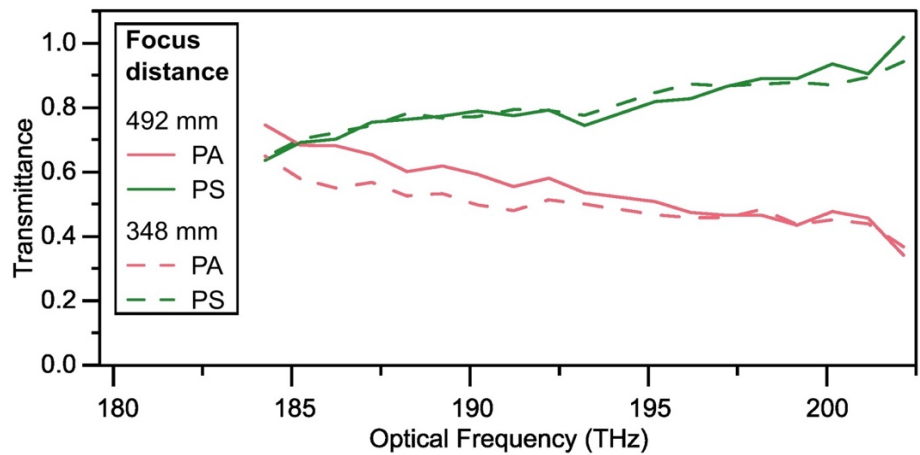

**Supplementary Figure 4. Plastic objects at different focus distances.** The transmittance spectra of four plastic particles (two of polyamide (PA) and two of polystyrene (PS)) placed at different distances along the beam axis. The good agreement between the transmittance spectra of the same species indicates that placement along the axis has no influence.

**Extended Discussion**
The application of digital holography in conjunction with frequency combs represents a recent development in the field. An initial report employed a frequency comb with a 250-MHz repetition frequency to achieve low-temporal-coherence holography for optical sectioning at discrete profile depths [32]. Subsequent research adopted a line-by-line sequential method, whereby individual lines from an electro-optic comb with a 10-GHz spacing were optically filtered for use in a sequence of digital holographic recordings with continuous-wave single-frequency laser lines at different frequencies [33]. In contrast, the work on dual-comb holography by Ref.[23] constituted the first proof-of-principle demonstration of simultaneous utilisation of all the lines from a comb - an electro-optic comb - for digital holography, with the reference beam derived from a second comb of slightly different line spacing. Our approach diverges significantly by leveraging a frequency comb with large line spacing, a microcomb, thereby enabling hyperspectral imaging in condensed matter systems and overcoming challenges for efficient, unambiguous phase reconstruction of objects ranging in size from micrometres to millimetres, even in the presence of steep surface discontinuities. For example, in Ref. [23], a one-cent coin was measured using combs with 100 lines and 1-GHz spacing, but the narrow line spacing and substantial low-frequency intensity noise—arising from camera sampling—resulted in a scatter of 7.9 μm and a measurement time of 94.5 s. In contrast, our method reduces the scatter for the same coin to 1.2 μm with a measurement time of 0.96 s, corresponding to a reduction in acquisition time by a factor exceeding 4,000 at equivalent signal-to-noise ratio, or an increase in signal-to-noise ratio by a factor of 65 at constant measurement time. The throughput achieved by Ref. [23] was $8.7 \times 10^4$ pixels s$^{-1}$, more than an order of magnitude lower than in our implementation.

In related work on coherent ranging with a microcomb, Ref. [9] reported a rate of $1.0 \times 10^5$ pixels s$^{-1}$ using sequential measurement of 30 comb lines, and noted that with a dedicated detector and analogue-to-digital converter for each comb line, the throughput could reach $3.0 \times 10^6$ pixels s$^{-1}$. In time-of-flight time-stretch ranging with a Fourier mode-locked laser and a supercontinuum source, Ref.[34] demonstrated a line-scan rate of $1.0 \times 10^6$ pixels s$^{-1}$ with a single photodetector.

Thus, in addition to yielding a substantially richer dataset encompassing the full complex-valued wavefront at any desired focal plane together with the hyperspectral response of the scene, our microcomb-based digital holography achieves a throughput at the state of the art, with considerable scope for further enhancement.

**References**

1. T. W. Hänsch. Nobel Lecture: Passion for precision. *Rev. Mod. Phys.* **78**, 1297-1309 (2006).
2. T. Udem, R. Holzwarth and T. Hänsch. Optical frequency metrology. *Nature* **416**, 233-237 (2002).
3. A. L. Gaeta, M. Lipson and T. J. Kippenberg. Photonic-chip-based frequency combs. *Nature Photonics* **13**, 158-169 (2019).
4. T. J. Kippenberg, A. L. Gaeta, M. Lipson and M. L. Gorodetsky. Dissipative Kerr solitons in optical microresonators. *Science* **361**, eaan8083 (2018).
5. L. Chang, S. Liu and J. E. Bowers. Integrated optical frequency comb technologies. *Nature Photonics* **16**, 95-108 (2022).
6. I. Kudelin, W. Groman, Q.-X. Ji, J. Guo, M. L. Kelleher, D. Lee, T. Nakamura, C. A. McLemore, P. Shirmohammadi, S. Hanifi, H. Cheng, N. Jin, L. Wu, S. Halladay, Y. Luo, Z. Dai, W. Jin, J. Bai, Y. Liu, W. Zhang, C. Xiang, L. Chang, V. Iltchenko, O. Miller, A. Matsko, S. M. Bowers, P. T. Rakich, J. C. Campbell, J. E. Bowers, K. J. Vahala, F. Quinlan and S. A. Diddams. Photonic chip-based low-noise microwave oscillator. *Nature* **627**, 534-539 (2024).
7. Y. Zhao, J. K. Jang, G. J. Beals, K. J. McNulty, X. Ji, Y. Okawachi, M. Lipson and A. L. Gaeta. All-optical frequency division on-chip using a single laser. *Nature* **627**, 546-552 (2024).
8. P. Trocha, M. Karpov, D. Ganin, M. H. P. Pfeiffer, A. Kordts, S. Wolf, J. Krockenberger, P. Marin-Palomo, C. Weimann, S. Randel, W. Freude, T. J. Kippenberg and C. Koos. Ultrafast optical ranging using microresonator soliton frequency combs. *Science* **359**, 887-891 (2018).
9. J. Riemensberger, A. Lukashchuk, M. Karpov, W. Weng, E. Lucas, J. Liu and T. J. Kippenberg. Massively parallel coherent laser ranging using a soliton microcomb. *Nature* **581**, 164-170 (2020).
10. M.-G. Suh and K. J. Vahala. Soliton microcomb range measurement. *Science* **359**, 884-887 (2018).
11. E. Obrzud, M. Rainer, A. Harutyunyan, M. H. Anderson, J. Liu, M. Geiselmann, B. Chazelas, S. Kundermann, S. Lecomte, M. Cecconi, A. Ghedina, E. Molinari, F. Pepe, F. Wildi, F. Bouchy, T. J. Kippenberg and T. Herr. A microphotonic astrocomb. *Nature Photonics* **13**, 31-35 (2019).
12. X. Xu, M. Tan, B. Corcoran, J. Wu, A. Boes, T. G. Nguyen, S. T. Chu, B. E. Little, D. G. Hicks, R. Morandotti, A. Mitchell and D. J. Moss. 11 TOPS photonic convolutional accelerator for optical neural networks. *Nature* **589**, 44-51 (2021).
13. J. Feldmann, N. Youngblood, M. Karpov, H. Gehring, X. Li, M. Stappers, M. Le Gallo, X. Fu, A. Lukashchuk, A. S. Raja, J. Liu, C. D. Wright, A. Sebastian, T. J. Kippenberg, W. H. P. Pernice and H. Bhaskaran. Parallel convolutional processing using an integrated photonic tensor core. *Nature* **589**, 52-58 (2021).
14. B. Corcoran, A. Mitchell, R. Morandotti, L. K. Oxenløwe and D. J. Moss. Optical microcombs for ultrahigh-bandwidth communications. *Nature Photonics* **19**, 451-462 (2025).
15. M. A. Guidry, D. M. Lukin, K. Y. Yang, R. Trivedi and J. Vučković. Quantum optics of soliton microcombs. *Nature Photonics* **16**, 52-58 (2022).
16. M. Kues, C. Reimer, J. M. Lukens, W. J. Munro, A. M. Weiner, D. J. Moss and R. Morandotti. Quantum optical microcombs. *Nature Photonics* **13**, 170-179 (2019).
17. K. L. Law and R. C. Thompson. Microplastics in the seas. *Science* **345**, 144-145 (2014).
18. M.-G. Suh, Q.-F. Yang, K. Y. Yang, X. Yi and K. J. Vahala. Microresonator soliton dual-comb spectroscopy. *Science* **354**, 600-603 (2016).
19. A. Dutt, C. Joshi, X. Ji, J. Cardenas, Y. Okawachi, K. Luke, A. L. Gaeta and M. Lipson. On-chip dual-comb source for spectroscopy. *Science Advances* **4**, e1701858 (2018).
20. M. Yu, Y. Okawachi, A. G. Griffith, N. Picqué, M. Lipson and A. L. Gaeta. Silicon-chip-based mid-infrared dual-comb spectroscopy. *Nat. Commun.* **9**, 1869 (2018).

21. T. Ideguchi, S. Holzner, B. Bernhardt, G. Guelachvili, N. Picqué and T. W. Hänsch. Coherent Raman spectro-imaging with laser frequency combs. *Nature* **502**, 355-358 (2013).
22. N. Picqué and T. W. Hänsch. Frequency comb spectroscopy. *Nature Photonics* **13**, 146-157 (2019).
23. E. Vicentini, Z. Wang, K. Van Gasse, T. W. Hänsch and N. Picqué. Dual-comb hyperspectral digital holography. *Nature Photonics* **15**, 890-894 (2021).
24. L. Wu, W. Xie, H.-J. Chen, K. Colburn, C. Xiang, L. Chang, W. Jin, J.-Y. Liu, Y. Yu, Y. Yamamoto, J. E. Bowers, M.-G. Suh and K. J. Vahala. AlGaAs soliton microcombs at room temperature. *Optics Letters* **48**, 3853-3856 (2023).
25. R. C. Thompson, W. Courtene-Jones, J. Boucher, S. Pahl, K. Raubenheimer and A. A. Koelmans. Twenty years of microplastic pollution research—what have we learned? *Science* **386**, eadl2746.
26. T. Latychevskaia and H.-W. Fink. Practical algorithms for simulation and reconstruction of digital in-line holograms. *Applied Optics* **54**, 2424-2434 (2015).
27. B. Xu, Z. Chen, T. W. Hänsch and N. Picqué. Near-ultraviolet photon-counting dual-comb spectroscopy. *Nature* **627**, 289-294 (2024).
28. X. Xue, Y. Xuan, Y. Liu, P.-H. Wang, S. Chen, J. Wang, D. E. Leaird, M. Qi and A. M. Weiner. Mode-locked dark pulse Kerr combs in normal-dispersion microresonators. *Nature Photonics* **9**, 594-600 (2015).
29. W. Liang, A. A. Savchenkov, V. S. Ilchenko, D. Eliyahu, D. Seidel, A. B. Matsko and L. Maleki. Generation of a coherent near-infrared Kerr frequency comb in a monolithic microresonator with normal GVD. *Optics Letters* **39**, 2920-2923 (2014).
30. E. Lucas, S.-P. Yu, T. C. Briles, D. R. Carlson and S. B. Papp. Tailoring microcombs with inverse-designed, meta-dispersion microresonators. *Nature Photonics* **17**, 943-950 (2023).
31. J. Gass, A. Dakoff and M. K. Kim. Phase imaging without $2\pi$ ambiguity by multiwavelength digital holography. *Optics Letters* **28**, 1141-1143 (2003).
32. K. Körner, G. Pedrini, I. Alexeenko, T. Steinmetz, R. Holzwarth and W. Osten. Short temporal coherence digital holography with a femtosecond frequency comb laser for multi-level optical sectioning. *Optics Express* **20**, 7237-7242 (2012).
33. E. Hase, Y. Tokizane, M. Yamagiwa, T. Minamikawa, H. Yamamoto, I. Morohashi and T. Yasui. Multicascade-linked synthetic-wavelength digital holography using a line-by-line spectral-shaped optical frequency comb. *Optics Express* **29**, 15772-15785 (2021).
34. Y. Jiang, S. Karpf and B. Jalali. Time-stretch LiDAR as a spectrally scanned time-of-flight ranging camera. *Nature Photonics* **14**, 14-18 (2020).